\documentclass[preprint]{revtex4-1} 
\usepackage{amssymb}
\usepackage{graphicx}
\usepackage{cases}

\begin{document}

\title{Coalescence of bubbles and drops in an outer fluid}

\author{Joseph D. Paulsen}
\email{paulsenj@umass.edu}
\author{R\'emi Carmigniani}
\author{Anerudh Kannan}
\author{Justin C. Burton}
\author{Sidney R. Nagel}
\affiliation{The James Franck and Enrico Fermi Institutes and The Department of Physics, The University of Chicago, 929 E 57th St., Chicago, IL 60637, USA}



\begin{abstract} 

When two liquid drops touch, a microscopic connecting liquid bridge forms and rapidly grows as the two drops merge into one. 
Whereas coalescence has been thoroughly studied when drops coalesce in vacuum or air, many important situations involve coalescence in a dense surrounding fluid, such as oil coalescence in brine. 
Here we study the merging of gas bubbles and liquid drops in an external fluid. 
Our data indicate that the flows occur over much larger length scales in the outer fluid than inside the drops themselves. 
Thus we find that the asymptotic early regime is always dominated by the viscosity of the drops, independent of the external fluid. 
A phase diagram showing the crossovers into the different possible late-time dynamics identifies a dimensionless number that signifies when the external viscosity can be important. 

\end{abstract}

\maketitle




During coalescence, two drops merge via the formation of an infinitesimal liquid bridge between them, which then expands to the size of the drops. 
The dynamics are driven by the Laplace pressure, which initially is singular due to the infinite curvature of the liquid interface at the point of contact. 
This coalescence singularity has been studied in the situation where the two drops coalesce in vacuum or air \cite{Hopper1984,Hopper1990,Herrera1995,Eggers1999,MenchacaRocha2001,Eggers2003,Wu2004,Bonn2005,Thoroddsen2005_2,Lee2006,Fezzaa2008,Case2008,Case2009,Paulsen2011,Paulsen2012,Sprittles2012,Baroudi2014,Paulsen2013}. 
These studies sought to understand the speed at which the neck radius, $r(t)$, expands as a function of $t$, the time since initial contact. 
Different dynamic regimes have been identified. 
However, in many natural settings \cite{Weertman1968,Navon1998,Espino2011} and industrial applications \cite{Evans1994,Eow2002,Ahn2006}, coalescence occurs inside a surrounding fluid that cannot simply be ignored. 

One would, in general, expect that the addition of an external fluid would lead to an even more complex phase diagram with a variety of regimes where different forces, from flows external as well as internal to the drops, compete to determine the dynamics. 
Even without a significant external fluid, the dynamics of drop coalescence is complicated and subtle due to the many length-scales over which flows can take place: the drop radius, $A$, the neck radius, $r$, the separation of the two drops at that radius, $r^{2}/A$, and the curvature at the neck minimum for viscous drops, $r^{3}/A^{2}$ \cite{Hopper1984,Eggers1999,Paulsen2012,Paulsen2013}. 

Some earlier experimental studies of two-fluid coalescence worked in a regime where the viscosity or density of the outer fluid was considered to be negligible for the dynamics \cite{Yao2005,Yokota2011}. 
One study that worked in the regime where the external viscosity was substantial \cite{Aryafar2008} reported that the larger of $\mu_{\text{in}}$ or $\mu_{\text{out}}$ (the viscosity inside or outside the drop, respectively) determines the coalescence rate when viscosity dominates over inertia. 
In contrast, a theory addressing the effect of an exterior fluid in the Stokes regime (where inertia can be completely neglected for the flows inside the drop) predicted that the outer fluid initially decreases the rate of neck expansion, $dr(t)/dt$, by a factor of 4, independent of the value of $\mu_{\text{out}}$ \cite{Eggers1999}. 
This theory, however, does not address late times or the case where the outer fluid dominates the dynamics as in the coalescence of bubbles. 
(Moreover, it was recently shown \cite{Paulsen2012,Paulsen2013} that the Stokes description can only apply when both the neck radius and the inner viscosity are sufficiently large.) 
Finally, it was predicted that in the two-fluid case, inertial forces are proportional to the sum \cite{Charles1960,Gilet2007} of the inner and outer fluid densities. 

Here, by identifying the different regimes of coalescence when an exterior fluid is present, we can sort out some of these different claims. 
In particular, we measure the scaling laws for $r(t)$ in the case of two bubbles or drops merging in an outer fluid that is dominated by either viscous or inertial forces. 
We also determine the crossovers between the different dynamic regimes. 
Our results show a clean separation of regimes that delineate when the viscosity or inertia either inside of, or external to, the drops will dominate the dynamics. 
Our analysis shows that the length scales in the external fluid are much larger than those inside the drops when $\mu_{\text{out}}\gg\mu_{\text{in}}$. 
This dramatically changes the competition between the different forces in the problem and leads to the appealing, although perhaps counter-intuitive, result that the inner fluid invariably dominates the asymptotic dynamics at small scales and early times. 
Finally, our work identifies a dimensionless number that indicates when the viscosity of the external fluid controls the dynamics.

\bigskip

{\bf \noindent RESULTS}

{\bf \noindent Experiment.}
In our experiments, we coalesce hemispherical drops (or bubbles) of radius $A$. 
We use combinations of water and glycerol to vary the viscosity of the drops. 
Salt is dissolved in the drops to make them electrically conductive. 
The drops or bubbles are submerged in silicone oils having a wide range of viscosity ($0.49$ mPa~s $<\mu_{\text{out}}<29000$ mPa~s) but little variation in density ($761$ kg m$^{-3}<\rho_{\text{out}}<976$ kg m$^{-3}$). 
The interfacial tension, $\gamma$, varies by less than a factor of $1.15$ in the two-fluid experiments for a fixed inner fluid and by a factor of $1.35$ for air bubbles in different silicone oils, allowing us to isolate the external viscosity. 
Additionally, by changing the glycerol and salt content of the inner fluid and by coalescing the drops in either silicone oil or air, we vary the surface tension between $23.5$ mN m$^{-1}$ and $82.5$ mN m$^{-1}$. 

In the absence of an external fluid, the dynamics is determined solely by the dimensionless neck radius, $r/A$, and the dimensionless Ohnesorge number, $Oh_{\text{in}}=\mu_{\text{in}}/\sqrt{\rho_{\text{in}}\gamma A}$, which is a ratio of viscous forces to inertial and surface tension forces. 
In that case, coalescence begins in the inertially-limited-viscous (ILV) regime where 
\begin{equation} 
r(t)/A = C_0 (\gamma/\mu_{\text{in}}A)t, 
\label{ILV_roft} 
\end{equation} 
\noindent where $C_0$ is a prefactor of order unity \cite{Paulsen2012,Paulsen2013}. 
In this regime, viscous stresses are dominant near the neck, but the large inertia of the drops (which must be pulled together by the small forces at the neck) prevents the purely viscous (Stokes) theory from applying \cite{Paulsen2012,Paulsen2013}. 
In our experiments, $Oh_{\text{in}}<1$, so in the absence of an outer fluid, the drops would begin their coalescence in the ILV regime and transition to a regime dominated by inertia at late times. 
For the outer fluid, we define $Oh_{\text{out}}=\mu_{\text{out}}/\sqrt{\rho_{\text{out}}\gamma A}$, which is varied from $0.0013$ to $210$ in our experiments. 


We use an ultrafast electrical method \cite{Burton2004,Case2008,Case2009,Paulsen2011,Paulsen2012,Paulsen2013} to probe the neck radius, $r(t)$. 
We complement the electrical measurements with high-speed imaging, which does not extend to early times due to the small neck height ($\sim r^2/A$) and the high curvature at the neck minimum ($\sim A^{2}/r^{3}$). 
For bubble coalescence, measurements are obtained only from imaging.

\medskip

\begin{figure}[bt] 
\centering
\begin{center}
\includegraphics[width=3.4in]{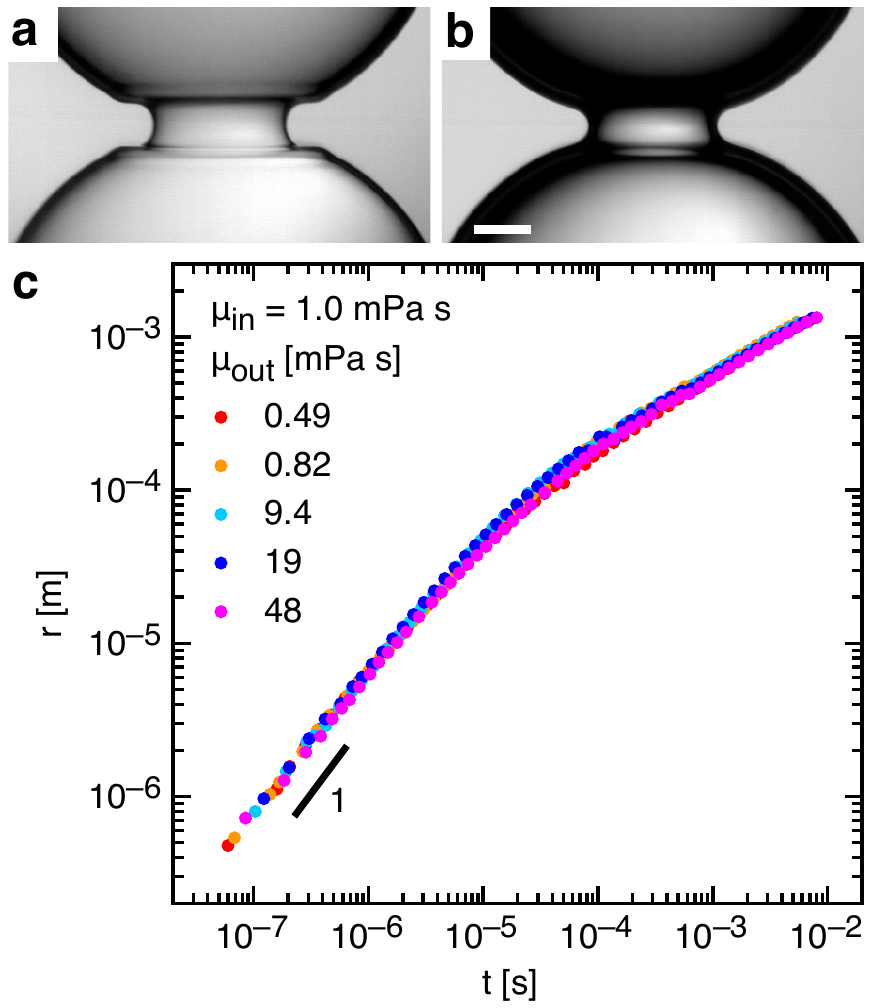}
\end{center}
\caption{
{\bf Salt water drops coalescing in silicone oils.} 
Salt water drops ($\mu_{\text{in}}=1.0$ mPa~s, $\rho=1070$ kg m$^{-3}$, $A=2$ mm) coalescing in silicone oil pictured $1.0$ ms after contact, with ({\bf a}) $\mu_{\text{out}}=0.49$ mPa~s, and ({\bf b}) $\mu_{\text{out}}=48$ mPa~s. 
Despite the large difference in $\mu_{\text{out}}$, the neck radii are nearly the same. 
The only difference is that capillary waves are visible in the less viscous outer fluid \cite{Thoroddsen2005_2}. 
Scale bar: $500$ $\mu$m. 
({\bf c}) Neck radius versus time for salt water drops coalescing in silicone oils of different viscosities. 
In these fluid combinations, $38$ mN m$^{-1}$ $<\gamma<40$ mN m$^{-1}$. 
The neck radius does not depend on the outer-fluid viscosity, even when it is 48 times more viscous than the liquid inside the drops. 
}
\label{Fig1}
\end{figure}

{\bf \noindent Salt water drops coalescing in outer fluids.}
Figure \ref{Fig1}a and b compares, at 1 ms after contact, salt water drops coalescing in silicone oils of viscosities varying by a factor of 100. 
The neck radii, $r(t)$, are essentially equal. 
Figure \ref{Fig1}c shows that $r(t)$ for salt-water drops is independent of the outer viscosity, even when $\mu_{\text{out}} \approx 50 \mu_{\text{in}}$. 

All the data are consistent with $r(t) \propto t$ at early times, as in equation \ref{ILV_roft} describing drop coalescence in air: 
the dynamics are dominated by the inner fluid despite the much more viscous surroundings.

\medskip

{\bf \noindent Coalescence of air bubbles in an outer fluid.} 
To understand the role of the outer fluid, we study the coalescence of air bubbles to approximate the limit where the interior fluid has negligible viscosity and density. 
In this case, there is no resistance to tangential flow at the drop interface so that the outer fluid can escape radially without significant axial velocity gradients over the small length scale $r^2/A$. 
Instead, the dominant gradients are in the radial direction over a length scale $L \approx r$. 
The driving force is the average Laplace pressure in the neck region, $\Delta P \approx \gamma A/r^2$. 
(Derivations of these choices for $L$ and $\Delta P$ are given in the Methods section.) 
With these choices for $L$ and $\Delta P$, we can estimate the velocity of the expanding bubble neck radius. 

When the inner fluid can be completely neglected and the external fluid is viscous, the viscous stress, $\mu_{\text{out}}(\partial u/\partial x)$, can be estimated by $\mu_{\text{out}}(U/L)=\mu_{\text{out}}(U/r)$, where $U=dr(t)/dt$ is the dominant velocity scale. 
Equating the viscous stress with the Laplace pressure, $\Delta P$, we get a differential equation that can be integrated to give: 
\begin{equation}
r(t)/A = C_1(\gamma /\mu_{\text{out}}A)^{1/2}t^{1/2}
= C_1 \left(\frac{t}{\tau_{\text{visc,out}}}\right)^{1/2}, 
\label{bubble_viscous_roft}
\end{equation}
\noindent where $C_1$ is a dimensionless prefactor and $\tau_{\text{visc,out}}=\mu_{\text{out}}A/\gamma$. 

Likewise, we can determine the dynamics when the inertial stress of the external fluid, $\rho_{\text{out}} U^2$, is dominant over its viscous stress. 
Setting $U=dr(t)/dt$ and equating the stress with $\Delta P$ leads to: 
\begin{equation}
r(t)/A = D_1 (\gamma/\rho_{\text{out}}A^3)^{1/4}t^{1/2}
= D_1 \left(\frac{t}{\tau_{\text{inert,out}}}\right)^{1/2}, 
\label{bubble_inertial_roft}
\end{equation}
\noindent where $D_1$ is also a dimensionless prefactor and $\tau_{\text{inert,out}}=\sqrt{\rho_{\text{out}} A^3/\gamma}$. 
(As noted previously 
\cite{Thoroddsen2005_1} and derived from energy-balance \cite{Czerski2011}, this last equation has the same form as for inertial coalescence of \emph{drops} in vacuum \cite{Eggers1999,MenchacaRocha2001,Eggers2003,Wu2004,Lee2006}, if $\rho_{\text{out}}$ is replaced by $\rho_{\text{in}}$.) 
Equations \ref{bubble_viscous_roft} and \ref{bubble_inertial_roft} indicate that the viscous and inertial regimes of bubble coalescence scale in the same way with only a difference in their characteristic time-scales.

\begin{figure}[bt] 
\centering
\begin{center}
\includegraphics[width=3.4in]{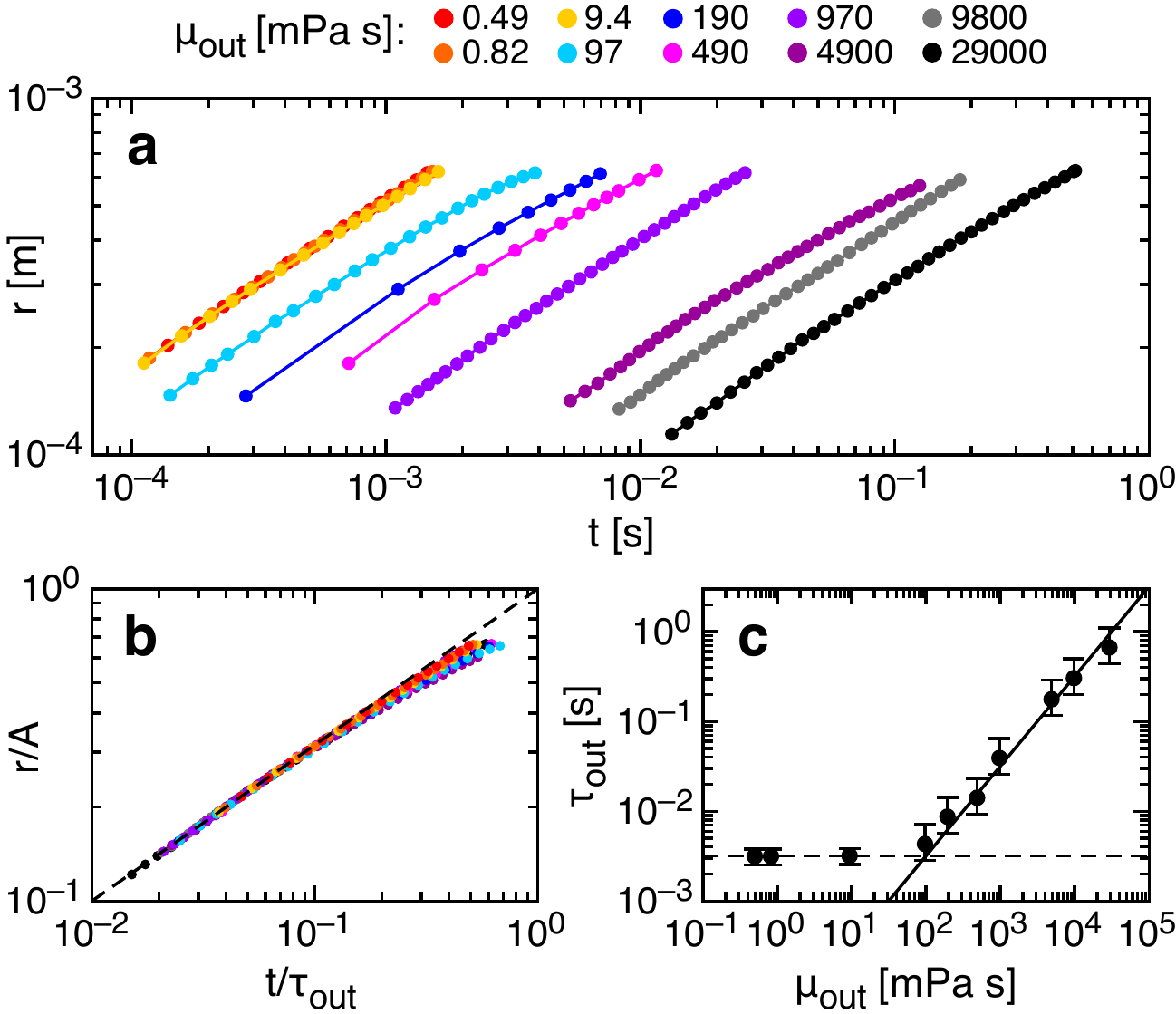}
\end{center}
\caption{
{\bf Air bubbles coalescing in silicone oils.} 
({\bf a}) Neck radius versus time measured optically. 
The outer-fluid viscosity is varied across a wide range; $\mu_{\text{out}}=0.49$ mPa~s to $29000$ mPa~s, while other parameters are held nearly constant ($\gamma=15.9$ to $21.5$ mN m$^{-1}$, $\rho=761$ to $976$ kg m$^{-3}$, $A=0.94$ mm). 
({\bf b}) Data rescaled by the drop radius, $A$, and a time-scale, $\tau_{\text{out}}$. 
The rescaled data follow $r(t)=(t/\tau_{\text{out}})^{1/2}$ (dashed line). 
The small departure at late times occurs when finite-size effects should become important as the neck radius approaches the size of the drops. 
({\bf c}) Coalescence time-scale, $\tau_{\text{out}}$, versus $\mu_{\text{out}}$ (error bars are from the fits to the data in ({\bf b})). 
At high-viscosity, $\tau_{\text{out}}$ is approximately equal to the viscous time-scale of the outer fluid (solid line: $\tau_{\text{out}}=0.72\tau_{\text{visc,out}}$ corresponding to $C_1=1.2$ in equation \ref{bubble_viscous_roft}). 
At low-viscosity, it is approximately given by the inertial time-scale of the outer fluid (dashed line: $\tau_{\text{out}}=0.51\tau_{\text{inert,out}}$ corresponding to $D_1=1.4$ in equation \ref{bubble_inertial_roft}). 
The lines intersect at $\mu_{\text{out}}=99$ mPa~s ($Oh_{\text{out}}=0.77$). 
}
\label{Fig2}
\end{figure}

To test these predictions, we show $r(t)$ versus $t$ in Fig.\ \ref{Fig2}a for air bubbles coalescing in silicone oils. 
All of the data have a similar slope. 
Thus we can collapse them onto the master curve shown in Fig.\ \ref{Fig2}b by rescaling the y-axis with the drop radius, $A$, and the x-axis with a measured time-scale, $\tau_{\text{out}}$, which we fit for each outer fluid to produce the best collapse. 
 
We plot $\tau_{\text{out}}$ versus $\mu_{\text{out}}$ in Fig.\ \ref{Fig2}c. 
There are clearly two distinct regimes. 
For high viscosities, $\tau_{\text{out}} \approx 0.72\tau_{\text{visc,out}}$, corresponding to $C_1=1.2$. 
For low viscosities, $\tau_{\text{out}} \approx 0.51\tau_{\text{inert,out}}$, corresponding to $D_1=1.4$. 
Both prefactors, $C_1$ and $D_1$, are of order unity as expected. 
In a separate analysis, we determine the scaling exponent by fitting the data to a power law: $r(t) \propto t^n$ and measure $n=0.55\pm0.09$ and $n=0.49\pm0.05$ at high- and low-viscosities respectively. 
Both are consistent with $n=1/2$. 
Thus, the data in both regimes are consistent with the predicted scaling laws, equations \ref{bubble_viscous_roft} and \ref{bubble_inertial_roft}.


\medskip

{\bf \noindent Competition between inner and outer fluids.} 
Returning to the two-fluid case, we now consider the competition between the stresses inside and outside the drops. 
As the ratio $\mu_{\text{in}}/\mu_{\text{out}}$ decreases, there must be a transition from the behavior observed in Fig.\ \ref{Fig1} (where inner flows dominate) to that seen in Fig.\ \ref{Fig2} (where the external fluid is most important).

\begin{figure*}[bt] 
\centering
\begin{center}
\includegraphics[width=5.3in]{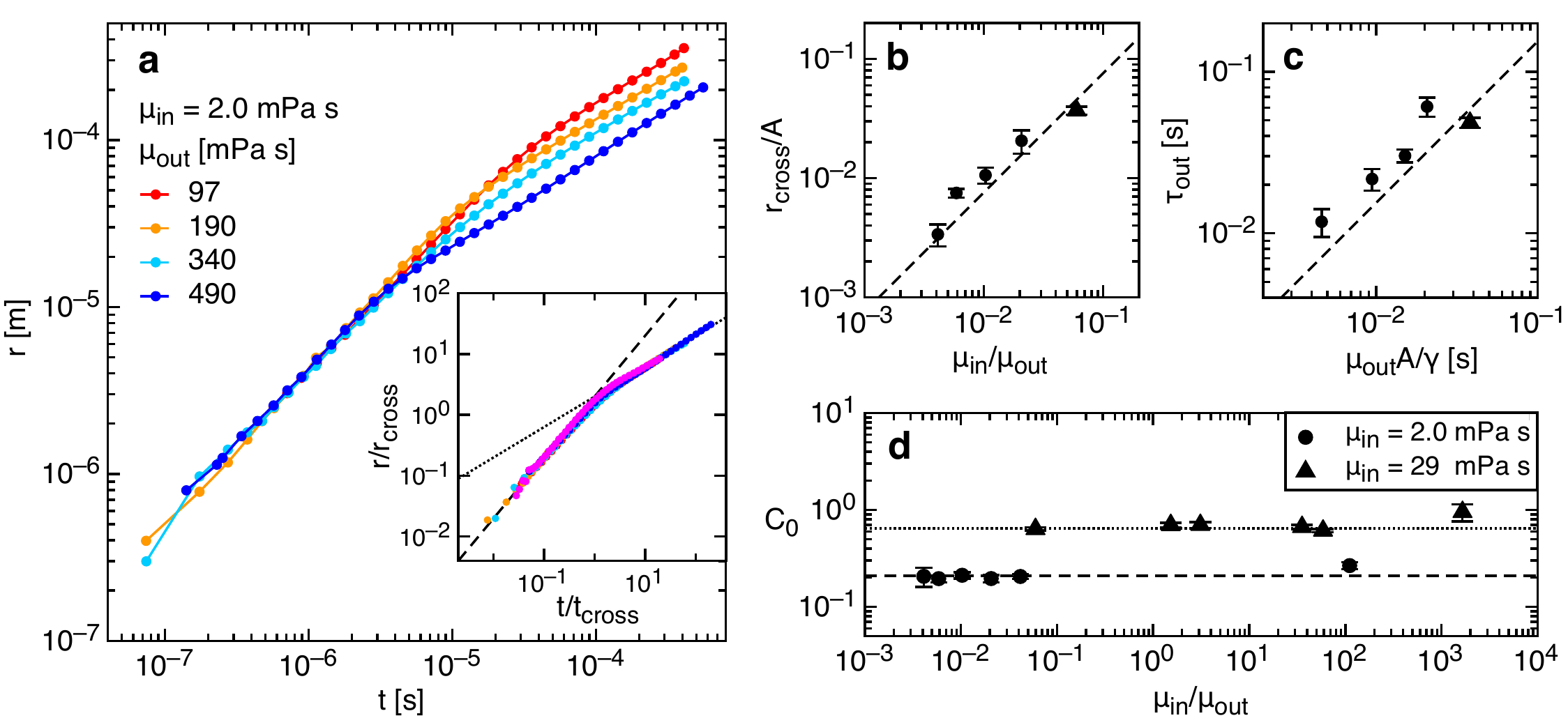}
\end{center}
\caption{
{\bf Inner-fluid to outer-fluid crossover.} 
({\bf a}) Neck radius versus time for salt water drops ($\mu_{\text{in}}=2.0$ mPa~s, $\rho=1200$ kg m$^{-3}$, $A=2$ mm) coalescing in silicone oils. 
Surface tension is roughly constant ($\gamma=41$ to $47$ mN m$^{-1}$). 
As $\mu_{\text{out}}$ is increased, the data departs from the linear scaling at earlier times. 
\emph{Inset:} The data rescaled by a crossover radius, $r_{\text{cross}}$, and crossover time, $t_{\text{cross}}$, to give the best collapse, including data with $\mu_{\text{in}}=29$ mPa~s and $\mu_{\text{out}}=490$ mPa~s (pink symbols). 
The dashed line has slope 1 and the dotted line has slope 1/2. 
({\bf b}) Inner-outer crossover radius, $r_{\text{cross}}$, divided by drop radius, $A$, versus viscosity ratio $\mu_{\text{in}}/\mu_{\text{out}}$ (circles: $\mu_{\text{in}}=2.0$ mPa~s, triangles: $\mu_{\text{in}}=29$ mPa~s). 
The data is well-described by $r_{\text{cross}}/A=0.76 \mu_{\text{in}}/\mu_{\text{out}}$ (dashed line) consistent with a crossover from an ILV regime to a regime dominated by the viscosity of the outer fluid. 
({\bf c}) $\tau_{\text{out}}$ versus $\mu_{\text{out}}A/\gamma$ at late times. 
The data follow $\tau_{\text{out}}=1.5 \mu_{\text{out}}A/\gamma$ (dashed line) indicating that the viscosity of the outer fluid dominates this regime. 
({\bf d}) Scaling prefactor, $C_0$, versus viscosity ratio, $\mu_{\text{in}}/\mu_{\text{out}}$. 
For fixed inner viscosity, the prefactor is independent of $\mu_{\text{out}}$ (shown by the horizontal lines). 
In ({\bf b}) and ({\bf c}), $\gamma=25.5$ to $47$ mN m$^{-1}$; in ({\bf d}), $\gamma=23.5$ to $82.5$ mN m$^{-1}$. 
In ({\bf b-d}), the error bars are from the fits to the $r(t)$ data. 
}
\label{Fig3}
\end{figure*}


In Fig.\ \ref{Fig3}a, we show data for $r(t)$ for salt-water drops coalescing in outer fluids of different viscosities. 
This is similar to Fig.\ \ref{Fig1}c but we have now extended the range to much smaller viscosity ratios, $\mu_{\text{in}}/\mu_{\text{out}}$. 
The early-time data is linear over the entire range, suggesting that the dynamics are still dominated by the inner fluid in the ILV regime. 
A fit to the data at later time gives: $r(t) \propto t^{0.54 \pm 0.03}$, which is consistent with what we see in bubble coalescence. 
Thus when $\mu_{\text{out}}\gg \mu_{\text{in}}$, a single coalescence event has a crossover from where the dominant flows are initially interior to where they are eventually exterior to the drops. 
The data can be collapsed onto a master curve if we rescale by a crossover time, $t_{\text{cross}}$, and crossover radius, $r_{\text{cross}}$, as shown in the inset. 

In Fig.\ \ref{Fig3}b, the dashed line shows that there is an approximately linear dependence of the crossover radius on the viscosity ratio: $r_{\text{cross}}/A \approx 0.76 \mu_{\text{in}}/\mu_{\text{out}}$. 
To reinforce that the late-time behavior is dominated by the outer fluid, 
Fig.\ \ref{Fig3}c shows $\tau_{\text{out}} \approx 1.5 \mu_{\text{out}}A/\gamma$, indicating that the outer-fluid viscosity indeed controls the late-time dynamics. 
Using equation \ref{bubble_viscous_roft}, we find $C_1=0.81$. 
The presence of an inner fluid has thus changed the prefactor, $C_1$, from what it was for bubbles. 
It has not, however, changed the dependence of $r(t)$ on time or on external viscosity. 

Finally, we test whether the outer fluid has any effect on the initial regime of drop coalescence. 
Fitting to equation \ref{ILV_roft}, Fig.\ \ref{Fig3}d shows the numerical prefactor, $C_0$, versus $\mu_{\text{in}}/\mu_{\text{out}}$. 
This prefactor is constant to within experimental error over a wide range of $\mu_{\text{out}}$ when $\mu_{\text{in}}$ is fixed. 
($C_0$ depends weakly on $\mu_{\text{in}}$, as was observed for drop coalescence in air \cite{Paulsen2011,Paulsen2013}.) 
We note that the points with the largest viscosity ratio, $\mu_{\text{in}}/\mu_{\text{out}}$, correspond to drop coalescence in air, where $\rho_{\text{out}}$ is $630$ to $810$ times smaller than in the rest of the data. 
These results indicate that the presence of the external fluid does not alter the early-time behavior---coalescence always starts in the ILV regime of equation \ref{ILV_roft}.

\medskip

{\bf \noindent Possible crossovers between the regimes.} 
We now consider the different possible crossovers that can exist as a pair of drops coalesce in an outer fluid. 
We do the most naive approximation and simply consider the crossovers between the four possible regimes outlined in Table \ref{Tab1}. 
To determine the crossover, we estimate the peak stress as a function of neck radius, for each regime. 
When the stresses in two regimes are equal, there will be a crossover from one regime to the other.

\setlength{\tabcolsep}{3.0pt}
\begin{table*}[bt]
\begin{tabular}{lllll}
\hline
\hline
Regime & Neck scaling & Stress scale & Crossover $r_{\text{cross}}/A$\\
\hline
Inertially-limited-viscous & $(\gamma/\mu_{\text{in}})t$ & $\mu_{\text{in}} (\frac{dr(t)}{dt}) A/r^2$ & \\
Outer-viscous & $(\gamma A/\mu_{\text{out}})^{1/2}t^{1/2}$ & $\mu_{\text{out}} (\frac{dr(t)}{dt})/r$ & $\frac{\mu_{\text{in}}}{\mu_{\text{out}}}$ (ILV to outer-viscous) \\
Inner-inertial & $(\gamma A/\rho_{\text{in}})^{1/4}t^{1/2}$ & $\rho_{\text{in}}(\frac{dr(t)}{dt})^2$ & $\mu_{\text{in}}/\sqrt{\rho_{\text{in}}\gamma A}$ (ILV to inner-inertial) \cite{Paulsen2011} \\
Outer-inertial & $(\gamma A/\rho_{\text{out}})^{1/4}t^{1/2}$ & $\rho_{\text{out}}(\frac{dr(t)}{dt})^2$ & $\mu_{\text{in}}/\sqrt{\rho_{\text{out}}\gamma A}$ (ILV to outer-inertial) \\ 
\hline
\hline
\end{tabular}
\caption{
{\bf Regimes of two-fluid coalescence for $Oh_{\text{in}}<1$.} 
For each regime, we list the scaling of the neck radius versus time, the dominant stress, and the dimensionless crossover radius $r_{\text{cross}}/A$, omitting dimensionless prefactors of order unity.
}
\label{Tab1}
\end{table*}

The ILV regime has the most rapidly diverging stress at early time (small $r$). 
Therefore in a continuum approximation, all coalescence must be asymptotically dominated by the dynamics within the drops. 
(Of course, if the scale where the inner viscosity dominates is below the size of an atom, then the ILV regime is cut off.) 
After starting in the ILV regime, the dynamics can transition into the outer-viscous, the inner-inertial, or the outer-inertial regimes. 
By equating stresses, we calculate the dimensionless neck radius, $r/A$, for each of these crossovers. 
We list these in Table \ref{Tab1}. 
An ILV to outer-viscous crossover should occur when $r/A\approx \mu_{\text{in}}/\mu_{\text{out}}$, consistent with our measurements in Fig.\ \ref{Fig3}b. 
We expect an ILV to inner-inertial crossover when $r/A\approx Oh_{\text{in}}$. 
This is the transition seen in Fig.\ \ref{Fig1}c and for drops coalescing in air \cite{Paulsen2011,Paulsen2013}. 
Finally, we predict that if $\rho_{\text{out}}$ is sufficiently large, an ILV to outer-inertial crossover is possible, when $r/A\approx \mu_{\text{in}}/\sqrt{\rho_{\text{out}}\gamma A}$. 
(This would occur outside of the range of our bubble coalescence experiments.) 

Crucially, we observe that the time dependance of the stresses in all regimes \emph{except} the ILV regime are identical---they all decay as $1/t$. 
(This comes from plugging $r(t)$ into the stress scale of each regime.) 
Therefore, once a crossover occurs out of the ILV regime into a second regime, coalescence continues in that regime until the drops have completely merged. 
This explains why the data in Fig.\ \ref{Fig1}c were completely independent of the value of $\mu_{\text{out}}$; for these fluid parameters, the drops transition from the ILV regime into an inertial regime. 
They remain in that inertial regime to the end and the external viscosity does not play a role. 
This also implies that bubbles coalescing in an outer fluid will not have a crossover between the outer-viscous and outer-inertial regimes as a function of time. 
Instead, the phase boundary between the outer-viscous and outer-inertial regimes is independent of $r(t)$ and is given by $Oh_{\text{out}}\approx 1$, consistent with our measurements in Fig.\ \ref{Fig2}c.

\medskip

\begin{figure}[bt] 
\centering
\begin{center}
\includegraphics[width=2.2in]{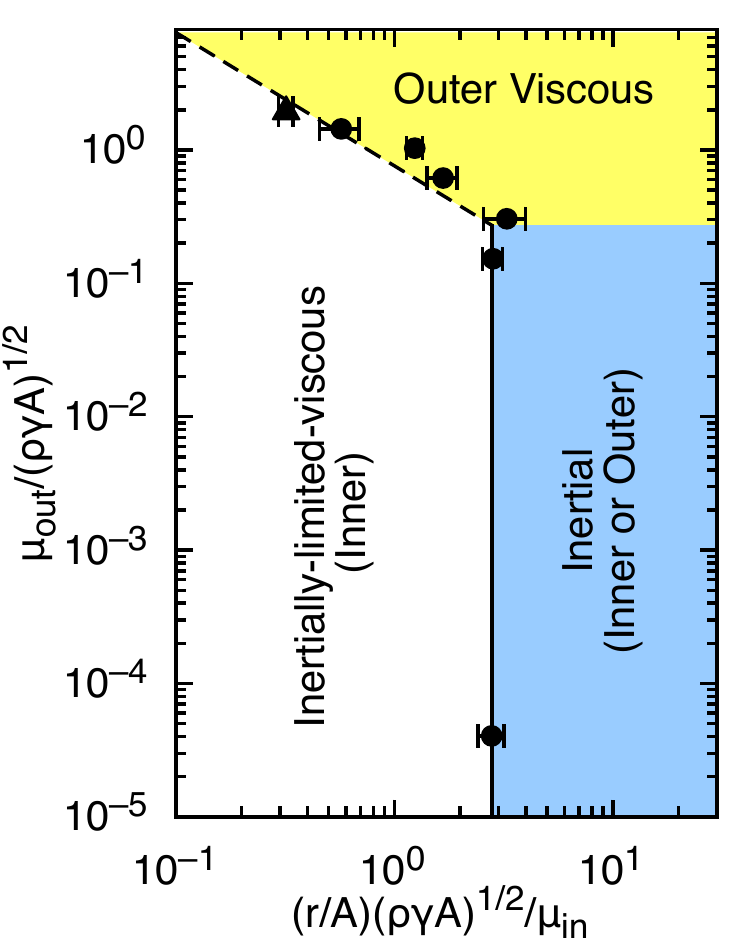}
\end{center}
\caption{
{\bf Two-fluid and bubble coalescence phase diagram for $Oh_{\text{in}}<1$.} 
Coalescence begins in the inertially-limited-viscous regime, where the neck radius grows independent of the outer-fluid viscosity. 
For $\mu_{\text{out}}/\sqrt{\rho\gamma A}<0.3$, inertia takes over at late times (solid line: $r/A=2.8 \mu_{\text{in}}/\sqrt{\rho\gamma A}$), where $\rho$ is the density of the more dense fluid. 
If instead $\mu_{\text{out}}/\sqrt{\rho\gamma A}>0.3$, then the outer-fluid viscosity dominates at late times (dashed line: $r/A= 0.76 \mu_{\text{in}}/\mu_{\text{out}}$). 
Symbols are measured crossovers from the data reported in Figs.\ \ref{Fig1} and \ref{Fig3}: (circles: $\mu_{\text{in}}=2.0$ mPa~s, triangles: $\mu_{\text{in}}=29$ mPa~s). 
For those data $\rho=\rho_{\text{in}}$. 
Surface tension ranges from $\gamma=25.5$ to $82.5$ mN m$^{-1}$. 
The point with the smallest $\mu_{\text{out}}/\sqrt{\rho\gamma A}$ is salt water coalescing in air. 
The error bars are from the fits to the $r(t)$ data. 
}
\label{Fig4}
\end{figure}

{\bf \noindent Two-fluid phase diagram.}
We assemble these results in a phase diagram for bubble and two-fluid coalescence, shown in Fig.\ \ref{Fig4}. 
Coalescence begins (at asymptotically early times) in the ILV regime where the outer fluid is unimportant, no matter how large its density or viscosity. 
In making the axes non-dimensional, an important dimensionless number emerges, given by $\mu_{\text{out}}/\sqrt{\rho\gamma A}$ (where $\rho$ is the higher of the two fluid densities). 
This number is determined from where the inertial stress (given by the inner or outer fluid) is equal to the viscous stress in the outer fluid. 
For $\mu_{\text{out}}/\sqrt{\rho\gamma A}<0.3$, inertia takes over at late times whereas if $\mu_{\text{out}}/\sqrt{\rho\gamma A}>0.3$, then the outer-fluid viscosity dominates at late times. 

This phase diagram implies that even for air bubbles coalescing in outer fluids, the viscosity of the inner fluid sets $dr(t)/dt$ at early times, which can therefore be very fast. 
For air bubbles coalescing in water with $A=2$ mm, equation \ref{bubble_inertial_roft} predicts that $dr(t)/dt$ exceeds the speed of sound in water for $r<0.25$ $\mu$m. 
This can produce shock waves in the water. 
Thus, compressibility effects will be important during the early moments of bubble coalescence. 

At very small neck radii, where the drop surfaces are very close to one another, van der Waals forces can become important and, in principle, affect the scaling results derived above. 
At worst, this could only affect our earliest electrical data, but not our bubble coalescence data, which does not probe to such small scales. 
Moreover, the effect of van der Waals forces will be mitigated because we expect the neck to form when the drops or bubbles are a finite distance apart. 
The presence of this gap will not change the expected scalings. 
(See Methods section.) 

\bigskip

{\bf \noindent DISCUSSION}

\noindent In summary, we have examined liquid drops with $Oh_{\text{in}}<1$ coalescing in an outer fluid. 
We showed that the outer fluid has a surprisingly small effect on the coalescence dynamics. 
Moreover, the inertially-limited-viscous regime is the asymptotic regime of liquid-drop coalescence, even in an outer fluid with significant density or viscosity. 
We expect the same to be true for $Oh_{\text{in}}>1$, for the simple reason that the force balance argument that identifies the ILV regime \cite{Paulsen2012} is only strengthened by having a second, ambient fluid with significant density. 
In that argument, the acceleration of the center-of-mass motion of a drop in the Stokes regime is compared with the forcing from surface tension which becomes arbitrarily small for small neck radius. 
When there is a surrounding fluid, the total mass that must be moved to bring the two drops together can only be larger than it is in vacuum. 
Therefore, we expect that the ILV regime should remain the asymptotic early-time regime for two-fluid coalescence, just as it is for the case with no external fluid. 
Further experiments are required to study the two-fluid case in the Stokes regime (which we expect to occur only at late times for $Oh_{\text{in}}>1$), where there is an analytic theory \cite{Eggers1999}. 

We note that in our scaling analysis for the two-fluid case, we have greatly simplified our picture by assuming that, at each point in time, one fluid can be completely ignored with respect to the dynamics of the other. 
In reality, the non-dominant fluid provides a perturbation that would affect the dimensionless prefactors of the crossovers and scaling laws, and the neck shape (as in ref.~\cite{Eggers1999}). 
Our data for drops and for bubbles coalescing inside a dominantly viscous external fluid show that the prefactor can change by a factor of $\approx 1.5$ but the scaling exponent is unaffected. 

For the case of air bubbles coalescing in an outer fluid, we have experimentally determined the growth dynamics. 
Our measurements are consistent with our scaling arguments wherein the exponent for the growth of the neck is identical in the inertial and viscous regimes. 
A full theory of bubble coalescence would give a more rigorous justification and could provide insight on the flows outside of the neck region. 

Finally, our work has identified a dimensionless number in two-fluid coalescence, $\mu_{\text{out}}/\sqrt{\rho\gamma A}$ (where $\rho$ is the larger of the two fluid densities), which may be used to predict whether the viscosity of the ambient fluid will ever be significant in the dynamics. 
This is just the Ohnesorge number for the outer fluid when $\rho_{\text{out}}>\rho_{\text{in}}$. 
However, if $\rho_{\text{in}}>\rho_{\text{out}}$, then it is a different dimensionless number. 
As we showed in the case of coalescing water drops, the outer fluid does not matter even if it is $\approx 50$ times more viscous than the water itself. 

\bigskip

{\bf \noindent METHODS}

{\bf \noindent Experiment.} We measure the neck radius versus time, $r(t)$, for drops or bubbles coalescing in an outer fluid. 
High-speed imaging was used for bubble coalescence and some of the two-fluid experiments; electrical measurements were performed on all of the two-fluid experiments. 
The methods are in good agreement where we obtained both types of data. 

In both methods, two hemispherical drops or bubbles of radius $A$ are formed on vertically aligned nozzles. 
The drops or bubbles are sufficiently small so that distortions due to gravity are minor. 
For the case of drops, we use combinations of water and glycerol to vary the interior viscosity, and we dissolve in salt (NaCl) to make them electrically conductive. 
The drops or bubbles are submerged in various silicone oils (Clearco Products) having a wide range of viscosity ($0.49$ mPa~s $<\mu_{\text{out}}<29000$ mPa~s) but small variation in density ($761$ kg m$^{-3}<\rho_{\text{out}}<976$ kg m$^{-3}$). 

To initiate coalescence, one drop or bubble is grown with a syringe pump at low speed so that the interfaces are undeformed when they touch. 
When the outer-fluid viscosity is large, we instead bring the drops or bubbles close together and hold them there until they coalesce (usually within 10 to 30 minutes). 
For drops, we monitor the deformation by measuring their capacitance immediately before the moment of contact, $t=0$. 
For bubbles, deformation is visible for high $\mu_{\text{out}}$, but it is smaller than the neck radii we measure. 
We record the resulting coalescence dynamics with a high-speed digital camera (Phantom series, Vision Research). 

In the electrical method \cite{Burton2004,Case2008,Case2009,Paulsen2011,Paulsen2012,Paulsen2013}, a high-frequency ($\geq 800$ kHz) low-amplitude ($\leq 1$ V) AC signal is applied across a known circuit element and across the drops as they coalesce. 
By varying the voltage and the frequency, we determined that the electric fields do not influence the coalescence dynamics of the expanding liquid neck \cite{Paulsen2013}. 
Sampling the output at high-speed, we follow ref.~\cite{Paulsen2011} to extract the complex impedance of the coalescing drops and convert it to a neck radius as a function of time: $r(t)$. 

Viscosities of the glycerol-NaCl-water mixtures were measured with glass capillary viscometers (Cannon-Fenske). 
Density was measured by weighing a known volume of fluid. 
We measured the interfacial tension, $\gamma$, for each combination of inner and outer fluids to within $\pm 1$ mN m$^{-1}$ by analyzing pictures of static pendant drops. 
For the fluid combinations used, $\gamma$ varied by less than a factor of $1.15$ for a fixed inner fluid. 
The values are given in the figure captions. 

We also measured the surface tension for each oil, as well as the viscosity and density of several oils, and the measurements were found to be consistent with the manufacturer product specifications. 


\medskip

{\bf \noindent Length scale for outer fluid flows.}
Here we argue that when bubbles are coalescing in an ambient fluid and the interior gas has negligible viscosity and density, then the radial flow gradients of the outer fluid are over a length scale comparable to the bubble neck radius, $r$. 

The gap between the bubbles at a radial distance $L$ from the neck (of radius $r$) is given to first order by $(r+L)^2/A$. 
Denoting the average radial velocity there as $v_{L}$, continuity for an incompressible outer fluid gives: $(2\pi r^3/A)(dr(t)/dt)=(2\pi (r+L)^3/A)v_{L}$. 
We wish to identify the length scale, $L$, for which $v_{L}$ decays to some small fraction, $1/N$, of the neck speed, $dr(t)/dt$. 
Setting $v_{L}=(1/N)(dr(t)/dt)$, we find: $L=r(N^{1/3}-1)\approx r$. 
(In two dimensions, $L=r(N^{1/2}-1)\approx r$.) 

\medskip

{\bf \noindent Laplace pressure scaling.}
The value of the Laplace pressure at the neck minimum is determined by the principal radii of curvature at that point. 
Depending on the coalescence regime, the dominant radius of curvature can have a different dependence on $r$. 
For drops coalescing in vacuum in the ILV regime \cite{Paulsen2012,Paulsen2013} and in the Stokes regime \cite{Hopper1984,Eggers1999,Paulsen2012,Paulsen2013} it will be of order $r^3/A^2$; for Stokes coalescence in an external fluid at early times \cite{Eggers1999} it will be of order $r^{3/2}/A^{1/2}$. 
Other regimes might produce other forms. 
However, the pressure and the flows are spread out in space, over either an axial scale $r^2/A$, or a radial scale $r$. 
Therefore, the driving force should be determined by a spatially averaged Laplace pressure, $\Delta P=2\gamma \overline{H}=\gamma(\overline{\kappa}_1+\overline{\kappa}_2)$, where $\overline{H}$ is the mean curvature, averaged over the entire neck region, and $\kappa_1, \kappa_2$ are the principal curvatures. 
Here we show that to leading order, $\overline{H}$ is set by the spacing between the drop interfaces, $r^2/A$, and is independent of the shape of the neck. 

We consider the drops to be spheres with radius $A$ and centers on the $z$-axis, touching at the origin, $(x,y,z)=(0,0,0)$. 
We compute the curvature in the $(x,z)$ plane first. 
The interfaces of the spherical drops are approximated to first order by $z=\pm x^2/(2A)$. 
The axisymmetric neck interface follows some function $f(z)$, which joins smoothly to the two drops at the points $(x_*,\pm x_*^2/(2A))$ with slopes $f'\equiv df/dz=\pm A/x_*$, where $x_* \approx r$. 
The line curvature of $f(z)$ is $\kappa_1=f''/(1+(f')^2)^{3/2}$. 
Averaging over the neck, we get $\overline{\kappa}_1 = -(A/x_*^2) \int_{-x_*}^{x_*} \kappa_1 dz = -(A/x_*^2) \int_{-x_*}^{x_*} f'' dz/(1+(f')^2)^{3/2} = -(A/x_*^2) f'/\sqrt{1+(f')^2}|_{-x_*}^{x_*}= -(2A/x_*^2)/\sqrt{1+(x_*/A)^2}= -(2A/x_*^2)(1-\frac{1}{2}(x_*/A)^2+\cdots)$. 
To leading order, $\overline{\kappa}_1= -2A/x_*^2\approx -A/r^2$. 
This curvature is also present in two-dimensional (2D) coalescence. 

The curvature of the neck in the $(x,y)$ plane is simply $\kappa_2=1/r$, which is an upper bound for the average value over the neck region, $\overline{\kappa}_2$. 
This curvature need only be considered in the force-balance at late times (and is absent in 2D coalescence). 

\medskip

{\bf \noindent Effect of small neck size.}
Our scaling predictions for $r(t)$ are for an idealized version of coalescence, corresponding to a neck of radius $r$ and height $r^2/A$ growing on two spheres of radius $A$. 
This is the same idealization used in refs.~\cite{Hopper1984,Hopper1990,Eggers1999}. 
However, we expect the neck to form when the drops or bubbles are a finite distance, $z_0$, apart so the neck height is instead given by $z_0+r^2/A$. 
When $r\ll \sqrt{z_0 A}$, the gap between the drops is approximately constant; later on, $r\gg \sqrt{z_0 A}$ and so $z_0\ll r^2/A$ can be ignored. 
(This gap was found to be $z_0=280^{+370}_{-160}$ nm for salt-water drops of radius $A=2$ mm coalescing in air \cite{Paulsen2013}, so in that case, $\sqrt{z_0 A}\approx 20$ $\mu$m.) 

Among the stresses listed in Table \ref{Tab1}, only the viscous stresses change for a finite gap, $z_0$, since the inertial stresses depend only on the fluid density and the neck speed. 
The peak viscous stress in the inner fluid would be: $\mu_{\text{in}} (dr(t)/dt)/(z_0+r^2/A)$. 
In the outer fluid, applying the argument for a constant-height gap gives, as before, a length-scale of $L\propto r$. 
In our experiments, the crossovers are all observed when $r>6$ $\mu$m (and our bubble coalescence data is for $r>100$ $\mu$m), and we find good agreement with our scaling arguments using the approximation $z_0+r^2/A\approx r^2/A$.

\medskip


\bigskip

{\bf Acknowledgements.}
We thank Osman Basaran, Efi Efrati, and Wendy Zhang for many enlightening discussions. 
We thank Irmgard Bischofberger and Andrzej Latka for measurements of fluid parameters of the silicone oils used. 
JDP gratefully acknowledges a Grainger Foundation Fellowship. 
This work was supported by NSF Grant DMR-1105145, NSF-MRSEC DMR-0820054, and NSF-PREM DMR-0934192. 

\medskip

{\bf Author contributions.}
JDP, RC, AK, JCB, and SRN designed the experiments and interpreted the results. 
JDP, RC, and AK performed the experiments. 
JDP and SRN wrote the manuscript with revisions from all of the authors. 

\medskip

{\bf Competing financial interests.} The authors declare that they have no competing financial interests.

\end{document}